\documentclass[11pt]{article}

\usepackage{graphics}   
\usepackage{graphicx}
\usepackage{endnotes}            
\usepackage{amsfonts}            
\usepackage{amssymb}             
\usepackage{amsthm}              
\usepackage{amsmath}            
\usepackage{mathrsfs}
\usepackage{algorithm}           
\usepackage{algorithmic}
\usepackage[letterpaper]{geometry}
\usepackage{rotating}            
\usepackage{tabularx}            

\usepackage{setspace}                
\usepackage{natbib}
\bibpunct{(}{)}{;}{a}{,}{;}

\begin{document}

\title{
Continuous Time Graph Processes with Known ERGM Equilibria: Contextual Review, Extensions, and Synthesis\thanks{This work was supported by NIH award 1R01GM144964-01 and NSF award SES-1826589. The author thanks Martina Morris, Michael Schweinberger, Chad Klumb, and Steve Goodreau for their input and helpful comments.}
}

\author{
Carter T. Butts\thanks{Departments of Sociology, Statistics, Computer Science, and EECS and Institute for Mathematical Behavioral Sciences, University of California Irvine; \texttt{buttsc@uci.edu}}
}
\date{2/12/23; to appear in \emph{Journal of Mathematical Sociology}, DOI 10.1080/0022250X.2023.2180001}
\maketitle

\begin{abstract}
Graph processes that unfold in continuous time are of obvious theoretical and practical interest.  Particularly useful are those whose long-term behavior converges to a graph distribution of known form.  Here, we review some of the conditions for such convergence, and provide examples of novel and/or known processes that do so.  These include subfamilies of the well-known stochastic actor oriented models, as well as continuum extensions of temporal and separable temporal exponential family random graph models.  We also comment on some related threads in the broader work on network dynamics, which provide additional context for the continuous time case.
\end{abstract}

\theoremstyle{plain}                        
\newtheorem{axiom}{Axiom}
\newtheorem{lemma}{Lemma}
\newtheorem{theorem}{Theorem}
\newtheorem{corollary}{Corollary}

\theoremstyle{definition}                 
\newtheorem{definition}{Definition}
\newtheorem{hypothesis}{Hypothesis}
\newtheorem{conjecture}{Conjecture}
\newtheorem{example}{Example}

\theoremstyle{remark}                    
\newtheorem{remark}{Remark}


Graph processes that unfold in continuous time are natural models for social network dynamics: able to directly represent changes in structure as they unfold (rather than, e.g. as snapshots at discrete intervals), such models not only offer the promise of capturing dynamics at high temporal resolution, but are also easily mapped to empirical data without the need to preselect a level of granularity with respect to which the dynamics are defined.  Although relatively few general frameworks of this type have been extensively studied, at least one (the stochastic actor-oriented models, or SAOMs) is arguably among the most successful and widely used families of models in the social sciences (see, e.g., \citet{snijders:sm:2001,steglich.et.al:sm:2010,burk.et.al:ijbd:2007,sijtsema.et.al:sd:2010,delahaye.et.al:jah:2011,weerman:c:2011,schaefer.kreager:asr:2020} among many others).  Work using other continuous time graph processes has also found applications both within \citep{koskinen.snijders:jspi:2007,koskinen.et.al:ns:2015,stadtfeld.et.al:sm:2017,hoffman.et.al:ns:2020} and beyond \citep{grazioli.et.al:jpcB:2019,yu.et.al:nsr:2020} the social sciences, suggesting the potential for further advances.

While some graph processes are non-ergodic, many of those used in network modeling lead to well-defined equilibrium graph distributions.  Of the latter, few have equilibria that are well-characterized.  For instance, a typical constant-rate SAOM of the type noted above will, in the large-time limit, converge from any initial condition to a well-defined distribution over the order-$N$ digraphs, but analytical expressions for such distributions are known only in trivial cases.  Although the behavior of such distributions can be explored through simulation, this proves costly in cases in which the natural dynamics of the system exhibit slow mixing.  More fundamentally, the inability to characterize the equilibrium behavior of such models limits our understanding of how structural dynamics lead to the observed incidence of structural forms, and may make it more difficult to identify and avoid model families with undesirable long-term behavior.

For some classes of continuous time graph processes, however, we can show not only the existence of an equilibrium graph distribution, but also the the form of that distribution.  These processes are of particular interest, since they allow us to directly link short-term dynamics with their long-term consequences.  Often, they are also inferentially tractable, with parameters that can be estimated from cross-sectional information supplemented by sometimes modest dynamic data (e.g., edge durations, or dwell times).  Further, it is frequently possible to study the long-term behavior of such graph processes by sampling directly from their equilibrium distributions (using, e.g., Markov Chain Monte Carlo (MCMC)), greatly economizing over explicitly simulated dynamics.  Here (and with minimal loss of generality), we focus on expressions of such equilibria in exponential-family random graph model form \citep{lusher.et.al:bk:2012}.  The exponential-family random graph models (ERGMs) are a highly general framework for describing distributions on sets of graphs \citep[see][for a review]{schweinberger.et.al:ss:2020}.  Indeed, the ERGMs are complete for finite support (i.e., for any distribution on a finite set of graphs, there exists an ERGM representation), and can represent broad and useful classes of distributions beyond this setting (e.g., for count-valued networks \citep{krivitsky:ejs:2012}).  The currently predominant use case is for unvalued networks on fixed vertex sets, which will be our focus here; however, many of the ideas discussed can be generalized beyond this case.

At present, ERGMs are employed primarily as models for single graph realizations (i.e., ``cross-sectional'' analysis), with no particular assumptions regarding associated dynamics. As a general language for specifying graph distributions, it is important to stress that the ERGM form itself has no \emph{inherent} substantive interpretation (though it has many extremely convenient mathematical, statistical, and computational properties.  However, ERGM forms arise ``naturally'' from a number of dynamic processes that are thought to be plausible accounts of structure formation in real systems, motivating a better understanding of the connection between ERGM families and stochastic processes.  Among other motivations: understanding ERGMs that arise from substantively plausible dynamics allows us to study the long-term behavior of those processes; dynamic interpretations of cross-sectional ERGMs can serve as useful conceptual devices for thinking about or understanding cross-sectional models; prior knowledge regarding network dynamics may help us better constrain the space of plausible cross-sectional models; in some cases, connections between dynamic processes and cross-sectional ERGMs may allow us to constrain the former from cross-sectional observations; and a deeper understanding of the connection between dynamic processes and cross-sectional ERGMs may help us design more efficient algorithms for simulation, parametric inference, or other computational tasks involving ERGMs.  This last is evidenced e.g. by the central role of quasi-time dynamics in the Markov chain Monte Carlo (MCMC) algorithms that are the backbone of ERGM computation.  The study of these relationships is thus of both practical and theoretical import.

While there are many types of graph processes that could be studied - and many connections between ERGMs and those processes - our focus here is specifically on \emph{continuous time processes with equilibrium graph distributions having a known ERGM form}.  From the above, we can immediately see that any continuous time process on a finite graph set that possesses an equilibrium distribution must trivially have an equilibrium in ERGM form; however, there is no obligation that this distribution be obvious from the dynamic specification, nor that it be easily specified.  Indeed, few cases where a continuous time process leads to a specified ERGM form are currently known.  In this paper, we review known families of continuous time graph processes with ERGM distributions, and also identify several additional families not previously reported in the literature.  In each case, we describe the family's transition rate structure and equilibrium behavior, demonstrating that each has the desired equilibrium using a uniform approach that aids in seeing similarities and differences across model families.  We also summarize a number of properties that distinguish these families from each other, and that may aid model selection in applied settings.

Given the broad range of problems that can examined within this space, we also note at the outset some questions that we do not address here.  With the exception of a few very brief remarks, we do not deal with the question of inference for model parameters, nor data-driven model selection (though we will comment on aspects of model behavior that may inform model selection on substantive grounds).  Likewise, we do not here discuss computational strategies for simulation of trajectories or equilibrium draws from graph processes, and we do not focus on specific examples of particular models in depth (as opposed to model classes).  All of these are important topics in their own right, but lie outside the scope of the present work.  It is hoped, however, that progress on these issues will be facilitated by a more complete catalog of candidate models to be studied.

The remainder of the paper is structured as follows.  Section~\ref{sec_notation} briefly reviews concepts and notation related to ERGMs and other formalisms that will be used throughout.  Section~\ref{sec_background} discusses more general background on related dynamic network processes that are not defined in continuous time and/or do not have known equilibria, but that provide useful context for understanding our processes of interest.  These are the focus of Section~\ref{sec_cont}, which discusses continuous time graph processes with ERGM equilibria, providing both novel cases and cases from the existing literature.  A comparative overview of the properties of these model families and related general issues are discussed in Section~\ref{sec_disc}, and Section~\ref{sec_conc} concludes the paper.

\section{Exponential Family Random Graph Models and Related Concepts} \label{sec_notation}

Although we shall for the most part introduce concepts and notation as we encounter them, it is useful to begin with a few notions that will be used throughout.  We start in particular with the exponential family random graph models.  Formally, let $G$ be a random graph on support $\mathbb{G}$.  A representation of the probability mass function of $G$ written as
\begin{equation}
\Pr(G=g|\theta,X) = \frac{\exp(\theta^T w(g,X)) h(g)}{\sum_{g'\in\mathbb{G}}\exp(\theta^T w(g',X)) h(g')} \label{eq_ergm}
\end{equation}
is said to be an \emph{ERGM form} for $G$, and any distribution written in such a form is generically referred to as an ERGM.  Here, $w:\mathbb{G},X \mapsto \mathbb{R}^k$ is a vector of \emph{sufficient statistics}, $\theta\in \mathbb{R}^k$ is a vector of \emph{model parameters}, $X$ is a covariate set, and $h: \mathbb{G} \mapsto \mathbb{R}_{\ge 0}$ is a \emph{reference measure} on $\mathbb{G}$.  Intuitively, $\theta$ biases (or ``tilts'') the distribution of $G$ with respect to the degrees of freedom indexed by $w$, relative to $h$; in particular, the conditional expectation of $w_i(G)$ is monotone in $\theta_i$.  It is common (but by no means necessary) to take $h(g) \propto I(g\in\mathbb{G})$, i.e. to specify ERGMs in terms of the \emph{counting measure} on $\mathbb{G}$.  In this case, the choice $\theta=0$ leads to the uniform distribution on $\mathbb{G}$, and the resulting family can be seen as an exponential tilting of the uniform random graphs (net of support constraints).  In the current development, we take $\mathbb{G}$ to be some set of unvalued graphs (directed or otherwise), although many of these ideas extend directly to the valued case.

For notational and conceptual simplicity, it is useful to observe that Eq.~\ref{eq_ergm} can be written as $\ln \Pr(G=g|\theta,X) = q(g) - \ln Z$, where $q(g) = \theta^T w(g,X) + \ln h(g)$ is a quantity we will call the \emph{graph potential} and $Z=\sum_{g'\in\mathbb{G}}\exp[q(g')+\ln h(g')]$ is the \emph{normalizing factor} or \emph{partition function}.  We notationally suppress the dependence of $q$ and $Z$ on $\theta$, $X$, $w$, and $h$ when there is no danger of confusion (i.e., we are interested only in the variation of $q$ with respect to $g$, and in $Z$ as a normalizer).  More explicit notation is invoked when needed.

While appropriate choices of $q$ can lead to ERGMs with very complex patterns of dependence among edges, it is also possible to specify ERGMs with no dependence among edges, or (respectively) dyads.  ERGMs in which all edges are independent are referred to as Bernoulli graphs (since edge presence/absence in such models is equivalent to independent Bernoulli trials).  Likewise, ERGMs on sets of directed graphs are referred to as Categorical graphs when there is no dependence between dyads, since the distribution of dyad states in such a model can be be treated as independent Categorical random variables.  In the undirected case, dyadic and edgewise independence are equivalent.  Neither Bernoulli nor Categorical graphs need be homogeneous (i.e., edge/dyad state probabilities may vary), and this is not assumed here unless explicitly stated.

In discussing network dynamics, it will often be useful to consider ``neighborhoods'' surrounding graphs in $\mathbb{G}$.  The most pragmatically important such neighborhoods are defined vis a vis the Hamming space of graphs, i.e. $\mathbb{G}$ together with the Hamming metric, $d$, such that $d(g,g')$ is the number of edge changes required to make $g$ into $g'$ (or vice versa).  Such edge state changes (adding or removing the $i,j$ edge) are often referred to in the literature as ``toggles.'' The set of all graphs $H$ such that $d(g,H)=k$ forms the Hamming sphere of radius $k$ about $g$, and the members of the radius-1 Hamming sphere about $g$ are said to be the (Hamming) neighbors of $g$.  We denote this set by $\mathcal{H}(g)$.  It is important to observe that typical choices of $\mathbb{G}$ in real models (e.g., the set of all order-$N$ graphs) are Hamming-connected, meaning that for any two graphs $g,g'\in \mathbb{G}$, there exists a sequence of graphs $g,\ldots,g' \in \mathbb{G}$ such that each graph is the Hamming neighbor of those adjacent to it in the sequence.  In such sets, it is possible to get from any graph in the support to any other graph, by applying a sequence of edge toggles.  In some cases, we will be interested in further narrowing down the set of Hamming neighbors, based on the type of toggle involved.  The subset of neighbors of $g$ reached by \emph{adding an edge} will be denoted $\mathcal{H}^{+}(g)$, while those reached by \emph{removing an edge} will be denoted by $\mathcal{H}^{-}(g)$; clearly, $\mathcal{H}(g)=\mathcal{H}^{+}(g) \cup \mathcal{H}^{-}(g)$.  We will also in some cases be interested in distinguishing between toggles that take $g$ into a graph of higher potential, versus a graph of lower potential.  We thus define $N^{+}(g) = \{g'\in \mathcal{H}(g): q(g')\ge q(g)\}$ and $N^{-}(g) = \{g'\in \mathcal{H}(g): q(g')< q(g)\}$ to be the set of ``uphill'' and ``downhill'' Hamming moves (respectively) with respect to the potential surface.

\section{Background: ERGMs and Network Dynamics} \label{sec_background}

Given our currently limited substantive understanding of network dynamics, the network modeling literature is surprisingly rich: models for network dynamics have been studied for a very long time \citep[since at least][]{katz.proctor:p:1959}, sometimes with similar or identical ideas being rediscovered or reinvented in new guises.  While our focus here is on the narrow question of continuous time network models that give rise to known equilibria, it is thus difficult to have a context for this work without considering either non-continuous time models with known equilibria (on the one hand), or continuous time models without known equilibria (on the other).  Of course, there is also a large set of models that are neither continuous nor that have known equilibria, but these are too far afield to be of interest.  While even these constraints leave considerable ground, we comment briefly on several topics that are particularly relevant to our focal concern.

\subsection{ERGMs from Pseudo-time and Quasi-time Dynamics}

As noted earlier, practical computation for ERGMs depends almost entirely on MCMC algorithms (introduced in the ERGM context by \citet{crouch.et.al:pres:1998} and \citet{snijders:joss:2002}, with subsequent elaborations and extensions by e.g. \citet{morris.et.al:jss:2008,wang.atchade:cissic:2014,byshkin.et.al:jsp:2016,butts:jms:2018}).  Broadly, current MCMC algorithms for ERGM simulation involve sequentially proposing single edge toggles, which are then accepted or rejected with a probability that depends on the current graph state.  As such, all such algorithms can be viewed as dynamic processes on $\mathbb{G}$ that operate in ``quasi-time:'' an imagined time dimension with no necessary connection to physical time, but that is relevant for the computational aspects of the respective algorithms (e.g. mixing time).  While quasi-time dynamics are usually encountered in computational settings, they can also be used to gain substantive insights.  For instance, \citet{morris.et.al:ajph:2009} use quasi-time dynamics scaled to roughly approximate plausible relationship durations to show how partnership concurrency can generate disparities in HIV prevalence, and \citet{butts:jms:2021} uses quasi-time dynamics along the order parameter of a phase transition in the edge/concurrent vertex model to provide an intuition on the mechanisms driving the transition, and how it might hypothetically unfold.  

It may be useful to distinguish such ``quasi-time'' dynamics - where there may be no simple correspondence between the timeline on which the dynamics unfold and physical time - from ``pseudo-time'' dynamics, in which some relationship is asserted, but not one of a cardinal nature.  For instance, many early ``dynamic network'' studies compared networks at two points in time, and simply examined how the later states were related to the earlier ones \citep[e.g.,][]{robins.pattison:jms:2001}.  Temporal ordering is present between earlier and later snapshots, but the model makes no other treatment of dynamics.  In another vein, dynamic processes much like those in the MCMC algorithms mentioned above can be motivated as stylized models of real network dynamics, but without a clear notion of pacing.  (See e.g., \citet{skvoretz:sn:1985,carley:asr:1991} for non-ERGM examples.)  In that case, it may be proposed that real dynamics are approximated by the network process, but the latter captures only the sequence and not the timing of events (``ordinal pseudo-time'').  Indeed, any Markov chain with a known ERGM distribution can be used in this way.  The lack of direct correspondence can even be seen as an advantage; for instance, \citet{butts:pres:2009,mele:wp:2010,mele:e:2017} introduce decision-theoretic models of network dynamics with this property (closely related to the SAOMs mentioned below), with the specific objective of interpreting cross-sectional network structure.  Because the equilibrium properties of the family hold under very loose conditions on the timing of decisions, it is not necessary to specify an exact timing model in order to obtain empirically useful results.  Of course, all such pseudo-time treatments are predictively limited by their inability to capture the pace of change, which by definition they do not represent, and more subtly to capture dynamic effects that depend on such pacing (e.g., high rates of turnover involving one set of relationships destabilizing another set of relationships).  This motivates models with explicit treatment of time.

\subsection{TERGMs and Other Discrete Time Models}

Perhaps the simplest way to capture physical time is to consider a graph series $G_0,G_1, \ldots$, indexed by some temporal variable $t_0,t_1,\ldots$.  Typically, the time periods between graphs are taken to be equal (though this can be relaxed).  This is a natural framework for treating network panel data, where we observe cross-sectional information on the state of a network at regular intervals in time.  In this context, discrete time is a coarsening of an underlying continuous time process, and models for such time series are viewed as approximating dynamics with respect to this coarsened process (which may omit mechanisms that operate on time scales much faster than the time scale of measurement).  It must also be noted that, in such settings, there are multiple ways in which the graph series may be defined relative to the underlying continuous time dynamics.  For instance, $G_i$ may reflect a snapshot of the state of a network at time $t_i$, or it may summarize the state of a network over the \emph{period} from $t_{i-1}$ to $t_i$ (or from $t_i$ to $t_{i+1}$, depending on definition).  Different representations may be natural in different settings (particularly when representations are chosen to correspond a particular type of measurement), and all have implications for modeling.

It should also be observed that some processes \emph{do} essentially occur in discrete time (or at least, are naturally treated in this way).  For instance, \citet{freeman.et.al:jsbs:1988} famously studied daily interactions among windsurfers on a California beach, over a one-month period.  Since the windsurfers convened on the beach each day (and were not continuously in residence), it is reasonable to treat the network as a genuinely discrete time process on a daily scale. 
Similar observations could be made regarding Baker's (\citeyear{baker:ajs:1984}) classic study of interactions among traders in a securities exchange, or any other study of networks that occur in settings that are regularly convened but that are not persistent.  

Whether they arise from coarsened continuous processes or from naturally episodic phenomena, distributions for graph series can be defined by positing an ERGM form for the conditional distribution of each graph within the series given those that have come before, i.e.,
\begin{equation}
\Pr(G_i=g_i | \theta,G_{<i}=g_{<i},X) = \mathrm{ERGM}(g_i|\theta,\{X\cup g_{<i}\}),
\end{equation}
where $g_{<i}$ refers to the time points prior to the $i$th, and $\mathrm{ERGM}(g|\theta,X)$ is the ERGM pmf evaluated at $g$ with parameters $\theta$ and covariates $X$ ($w$ and $h$ left tacit for brevity).  Models of this type are said to be \emph{temporal ERGMs} or TERGMs \citep{robins.pattison:jms:2001,hanneke.xing:ch:2007,desmarais.cranmer:phyA:2012}, and are the network analog of VAR models in classical time series analysis \citep{lutkepohl:bk:1993}.  All points made earlier regarding the generality of the ERGM representation apply here, although it must be borne in mind that if $s$, $h$, and/or $\theta$ are restricted to be time-homogeneous, the completeness of ERGMs for single graphs does not generalize to series thereof.  Even so, the TERGM framework can accommodate extremely general classes of discrete-time dynamics.  Partly because of this generality, most work using TERGMs focuses on sub-classes that are simpler to work with.  Common simplifications including imposing either first-order or $k$-th order Markov structure on past-dependence (so that present graph states are conditionally independent of states more than some $k$ steps removed in time), or simplifications on dependence among edges in the present.  Imposing complete edgewise independence in the present (given the past) typically leads to models with a simple regression structure, which are sometimes called dynamic network regression (DNR) families \citep{almquist.butts:pa:2013}.  Another strategy is to combine both ideas, using the previous time point to classify edge variables into those that are ``edge present'' and those that are ``edge absent'' and allowing conditional dependence only among edge variables in the same class.  This allows for distinct processes to be specified for the formation of new edges versus the persistence of existing ones, at the price of requiring conditional independence of formation and dissolution.  Families of this type are called \emph{separable TERGMs} or STERGMs \citep{krivitsky.handcock:jrssB:2014}, and are especially useful in systems for which the mechanisms governing the emergence of new relationships differ greatly from those leading to the cessation of existing relationships.

When employed as approximations to a continuous time process, it must be noted that the degree of conditional dependence among edges in an approximating TERGM depends on the relationship between the timescale of the TERGM process and the timescale of the underlying network dynamics on which it is based.  Observing that the past is a covariate to the present, it is intuitively obvious that (subject to some mild regularity conditions\footnote{E.g., we need the change rate to not go to infinity in finite time, so that the system cannot ``ergotize'' out from under us.}) $G_i$ will be increasingly determined by $G_{<i}$ as $t_{i}-t_{i-1} \to 0$, and the magnitude of approximation error resulting from omitting dependence in $G_i|G_{<i}$ will decline (if only because the entropy of $G_i|G_{<i}$ will itself decline).  Likewise, under similar regularity conditions,\footnote{We also need non-simultaneous edge state transitions, i.e., all non-vanishing transition rates are on unit moves in the Hamming space of the support.} we can appreciate that as $t_{i}-t_{i-1} \to 0$, all observations will contain either zero or one edge state change, and dependence is impossible.  This ``thin slicing'' limit can be used to motivate the use of DNR families for networks that are frequently measured.  As \citet{lerner.et.al:jmp:2013} have observed, the reverse of this is also true as measurement intervals grow: for typical processes, edge variables become increasingly dependent, and the DNR (and eventually, STERGM) approximations become poor.  Another implication of this phenomenon is that the magnitude of dependence in an approximating TERGM depends on the measurement interval, and not simply the process under study.  Adapting TERGMs estimated at one timescale to model dynamics on an alternative timescale hence requires non-trivial adjustments \citep{gibson.butts:sn:2023}.  Whether or not one regards this fact with displeasure, it is a universal truth that coarsened models differ depending on the degree of coarsening, and this is not unique to TERGMs.

In general, it should be observed that TERGMs need not have a well-defined equilibrium distribution (they can, e.g., be periodic, or if forced via $X$, undergo other types of continually fluctuating dynamics), nor are they necessarily intended to represent networks that are in equilibrium.  Further, even when TERGMs \emph{are} ergodic, we do not in general have a simple means of knowing the ERGM representation of their equilibrium distribution.  Other than trivial cases, or the pseudo-dynamic examples mentioned above, the family most studied in this regard is the family of STERGMs with constant dissolution rates \citep{carnegie.et.al:jcgs:2015}.  For this family, the equilibrium can be approximated by an ERGM with the statistics of the formation submodel and parameters that are a simple function of both submodels \citep{klumb.et.al:wp:2022}.  Importantly, this allows for STERGM parameters to be estimated from a single cross-sectional observation, provided that information on mean tie duration is also available; since, further, ERGMs depend only on their sufficient statistics, this allows some STERGM families to be fit to sampled egocentric data with duration measurements (making them feasible for use in large populations) \citep{krivitsky:tr:2012}.  This special case is obviously of particular practical importance, although many other TERGMs may also have interesting and useful equilibrium behavior that is so far unstudied.

\subsection{REMs, SAOMs, and Other Continuous Time Frameworks}

While not as common as discrete time approaches, various continuous time frameworks for network and/or social interaction dynamics have been proposed.  Most (the exceptions to which are our main point of interest here) are not intended to produce stable equilibrium behavior, and/or have equilibrium properties that are not in general known.  Only a few examples are mentioned here.  Continuous time models of a non-statistical sort are well-known in the agent-based modeling field \cite[e.g.][]{hummon.fararo:sn:1995,borgers.sarin:jet:1997,lawson.park:jasss:2000}, where they have traditionally been constructed with a view either to ensuring well-orderedness of social dynamics, or studying pacing in events.  As discussed below, there have been many proposals for continuous time models of edge state change in networks from a dyadic point of view, leading to fairly tractable models without dependence (but possibly with heterogeneity); these go back at least to the 1950s, and it seems likely that they have been rediscovered more than once.  A more significant development from the present point of view was the creation of the stochastic actor-oriented models (SAOMs) \citep{snijders:jms:1996,snijders.vanduijn:ch:1997,snijders:sm:2001}, a relatively general framework that combines an agent-based model of network dynamics with a statistical framework that allows for mechanistic drivers of interaction behavior to be inferred from cross-sectional data.  We will have more to say about a special case of stationary SAOMs with known ERGM distributions in Section~\ref{sec_cont}, but here merely sketch some salient characteristics.  SAOMs posit that the vertices of the network being modeled act as agents, who (in the standard formulation) have unilateral control over their outgoing edges.  These agents have utility functions over the states of the network (which they are assumed to accurately perceive), and when given opportunities to do so, will attempt to alter the state of a single edge so as to maximize their utility (myopically) for the resulting network; the choice set is taken to be the set of all unilateral edge changes, decisions must be made asynchronously, and neither forward nor backward looking behavior is assumed (i.e., agents neither remember past states, nor anticipate future responses of others to their own actions).  Choices are made via a multinomial logit, and are considered instantaneous.  Opportunities for such choices to be made are determined by a separate rate function (as Poissonian events), which is also part of the model specification; this can be a simple constant, or can be assumed to vary in one or another way with either the current state of the network or agent characteristics (or both).  While the generative process of the SAOM is a continuous time framework, it is worth noting that the \emph{inferential} uses of SAOMs are nearly always based on sequential cross-sectional observations (network panel data), and this has been the almost exclusive focus of work in this area \citep[see e.g., ][]{snijders:sm:2001,schweinberger.snijders:csda:2007,snijders.et.al:aas:2010,niezink.snijders:aas:2017}.  The continuous network dynamics are hence assumed to be fully latent, and only episodic snapshots are observed; changes of the network between snapshots are used to infer the drivers of dynamics (i.e., rate and utility functions, in the typical case).  In general, SAOMs are not assumed to be stationary (it is very common e.g. to assume time-varying effects that are presumed to be exogenously driven), and where equilibria exist, their forms are rarely known (see e.g. \citet{block.et.al:smr:2019} for a simulation-based investigation).  An important exception will be discussed below.

When continuous time dynamics can be observed directly (e.g., event history data), many of the inferential challenges of the SAOMs disappear, and it is possible to specify and perform inference for processes with many fewer constraints.  The relational event models (REMs) are one framework of this type \citep{butts:sm:2008}.  \emph{Relational events} are discrete events involving interactions between actors and/or actors and their environment, which can be effectively treated as instantaneous relative to the underlying dynamics of the system as a whole.  A REM consists of a population of such events, together with a specification for their hazards over time; it is very common for such models to assume locally Poissonian events with piecewise constant hazards, although this is not necessary.  (As with Cox models \citep{cox:jrssB:1972}, it is also possible to propose REMs with a non-parametrically specified and time-varying baseline hazard \citep{perry.wolfe:jrssB:2013}, though this limits their predictive value.)  Typically, REM specifications include interdependence among events, and may or may not be stationary; since inference for REMs is usually performed using observed event histories, stationarity is rarely necessary (and may not be desired).  Although REMs can be interpreted in behavioral terms \citep{butts:sm:2008,butts:sm:2017}, they need not be.  While REMs are usually employed to model instantaneous interaction patterns - and not, hence, networks - it is possible to model network dynamics in REM terms by positing two classes of events (formation and dissolution), and treating edges as the gaps between formation and dissolution events.  This approach has been leveraged e.g. by \citet{stadtfeld:bk:2012} to build SAOM-like models for use when event histories of network change are observable, allowing many constraints of the standard SAOMs to be relaxed.  It is noteworthy that this SAOM-like parameterization scheme \citep[the DyNAMs:][]{stadtfeld.et.al:sm:2017} can be translated into the scheme proposed by \citet{butts:sm:2008}, and vice versa; thus, the two are simply different ways of describing relational event processes.  However, the mapping between parameterizations is non-trivial, and each can have practical advantages in particular situations.  REM-like schemes have also been explored for graphs such as citation networks that grow entirely via formation, with edge addition treated as an instantaneous process that occurs when new nodes are added to the graph \citep{vu.et.al:icml:2011,vu.et.al:nips:2011}.    

\subsection{ERGMs as Physical Equilibria}

One last topic of relevance to our focal subject is the motivation of ERGMs as \emph{statistical mechanical} (as opposed to statistical) models.  In this interpretation, one views the focal graph, $G$, as a \emph{microstate} from an ensemble thereof, in energetic exchange with a thermal reservoir and observed at a \emph{random time.}  The detailed dynamics that give rise to an observation $g_t$ of $G$ are unobserved; however, in equilibrium, the probability of such an observation is
\begin{equation}
\Pr(G_t=g_t|\mathscr{H},T) = \frac{\exp\left[\tfrac{-\mathscr{H}(g_t)}{k_B T}+\sigma(g_t)\right]}{Z(\mathscr{H},T,\sigma)}, \label{eq_phys}
\end{equation}
where $\mathscr{H}$ is the Hamiltonian of the system (expressing the total energy associated with its topological degrees of freedom), $T$ is the temperature, $k_B$ is Boltzmann's constant, $\sigma$ is the entropy of the graph microstate, and $Z$ is the partition function.  If we write $\mathscr{H}$ with respect to a basis $w$ of functions on $\mathbb{G}$, we then have
\begin{equation}
\Pr(G_t=g_t|\phi,T) = \frac{\exp\left[\tfrac{-1}{k_B T}\phi^T w(g_t) + \sigma(g_t)\right]}{Z(\phi,T,\sigma)}, \label{eq_netham}
\end{equation}
which is plainly an ERGM with $\theta=\tfrac{-1}{k_B T}\phi$, and $h(g_t)=\exp(\sigma(g_t))$.  Aside from its immediate motivation in physical applications of ERGMs, this representation also provides insights into the behavior of ERGMs in other settings.  For instance, within the statistical mechanical picture, $\theta$ represents the effective generalized forces (relative to $k_B T$) driving the system, with $w$ being the degrees of freedom on which those forces act.  The reference measure can also be seen a generalized \emph{multiplicity,} representing (roughly, and sometimes exactly) the number of ways that a given graph can be generated by the unmodeled generative processes that create the network.  This makes explicit the role of unmodeled degrees of freedom in driving equilibrium behavior, and motivates greater theoretical attention to the oft-neglected reference measure.  (It also shows that there are conceptual differences between true parameter offsets and changes of reference measure, since one represents an energetic contribution while the other represents an entropic contribution.  In systems with a well-defined temperature, these behave very differently.)  Versions of this interpretation have (with varying degrees of specificity and attention to physical detail) been used by a number of researchers as tools or metaphors to study or explain ERGM behavior \citep[e.g][]{haggstrom.jonasson:jap:1999,park.newman:prevE:2004,robins.et.al:ajs:2005,radin.yin:aap:2013,butts:jms:2021}, and by others as physical models in their own right \citep{grazioli.et.al:jpcB:2019,yu.et.al:nsr:2020,diessner.et.al:jpcB:2023}.

From the standpoint of the present work, it is clear that the statistical mechanical interpretation of ERGMs treats them as arising from a latent continuous time process, and provides important insights into how this process translates into the equilibrium graph distribution.  As with quasi-time dynamics, thinking of ERGMs in terms of a physical equilibrium may be helpful in understanding model behavior (e.g., interpreting $\theta$s as forces).  At the same time, this approach does not by itself fix dynamics, and additional assumptions must be made for this purpose.  Below, we will examine one such scheme, which provides a useful mechanistic interpretation of the longitudinal ERGMs of \citet{koskinen.snijders:jspi:2007}.

\section{Continuous Time Processes with Known Graph Equilibria} \label{sec_cont}

With the foregoing as context, we now turn to the investigation of continuous time graph processes with well-defined equilibrium behavior, where that behavior can be expressed in terms of a graph distribution in ERGM form.  For compactness, we refer to these as \emph{ERGM generating processes}.  In what follows, we consider the following setup.  We assume the presence of a stochastic process, $S$, on state space $\mathbb{S}$, indexed by a temporal variable $t\in \mathbb{R}$ (such that $S_t$ is a random variable on support $\mathbb{S}$).  Taking $\mathbb{G}$ as before to be a graph set, we define the continuous extension of our random graph $G$ by the mapping $G: \mathbb{S} \mapsto \mathbb{G}$.  Thus, we may naturally define $G_t \equiv G(S_t)$ to be the graph associated with the state of $S$ at time $t$.  Note that, while it is possible to take $\mathbb{S}=\mathbb{G}$ (which indeed leads to a class of Markovian models), this is not always assumed; for instance, systems with memory will require a more elaborate state space.  For notational convenience, we will let realizations of $S$ be treated as time indexed sets, such that e.g. $s_t \in S$ can be read as ``the history $s$ of $S$ was at state $s_t$ at time $t$.''  We will also restrict attention to countable $\mathbb{S}$.

It is convenient to endow $\mathbb{S}$ with a topology, $T$, i.e. a directed graph representing allowable transitions; specifically, for all $s,s' \in \mathbb{S}$, if $s_t,s'_{t+\delta} \in S$ and $(s,s') \not\in T$, then there exists $t'\in (t,t_\delta)$ and $s''\in \mathbb{S}$ such that $(s,s'')\in T$ and $s''_{t'}\in S$, and there exists $t''\in (t,t_\delta)$ and $s'''\in \mathbb{S}$ such that $(s''',s')\in T$ and $s'''_{t''} \in S$.   We also define an \emph{instantaneous rate structure,} $R$, such that $R_{ss'}$ is the transition rate from state $s$ to state $s'$ for $s\neq s'$, and $R_{ss}=-\sum_{s'\in \mathbb{S}\setminus s} R_{ss'}$.  (Here, we treat memory or time-dependent covariates as folded into the state space, so $R$ can be treated as fixed.)  Clearly, $R_{ij}>0$ only if $(i,j)\in T$.  At present, we will confine ourselves to relatively well-behaved (Poissonian) processes that are characterized by the condition that, for all $s,s' \in \mathbb{S}$ and time $t$,
\begin{equation}
\lim_{\delta \to 0} \Pr(S_{t+\delta}=s'|S_t=s) = I(s=s') + \delta R_{ss'} + o(\delta)
\end{equation}
where $I$ is an indicator of its argument, and $o(\delta)$ are arbitrary terms that go to zero faster than $\delta$.  Equivalently, for such processes, 
\begin{enumerate}
\item For all $s\in \mathbb{S}$, $\sum_{s'\in\mathbb{S}} R_{ss'}=0$;
\item For all time $t$, the infimum $\delta>0$ such that $S_{t+\delta}\neq S_t$ is exponentially distributed with rate parameter $\sum_{s\in \mathbb{S}} R_{S_t s}$; and
\item For arbitrary time $t$, let $\delta>0$ be the infimum $\delta$ such that $S_{t+\delta}\neq S_t$; then $\Pr(S_{t+\delta}=s)=R_{S_t s}/\left[\sum_{s'\in \mathbb{S}\setminus S_t}R_{S_t s'}\right]$.
\end{enumerate}
Other characterizations are also possible.  Processes with these characteristics are of course \emph{continuous time Markov chains} (CTMCs), and have many salubrious properties \citep[see e.g.][]{grimmett.stirzaker:bk:1992}.  We note that it is not only possible to create stochastic processes with much more elaborate structure, but that examples of such processes do exist in the network literature (e.g., non-stationary DyNAMs); however, we are not aware of any examples of such processes leading to known ERGM forms, and thus not consider these cases here.

\subsection{Conditions for Convergence}

Although there is no known recipe that allows for specification of an equilibrium graph distribution from $S$ in all cases, there are sufficient conditions that can in some cases be used to establish that an equilibrium exists (and that can be exploited in seeking its form).  We here review some of these, again working within the above CTMC framework.  The underlying results are well-known, and can be found in any standard stochastic process text \citep[e.g][]{grimmett.stirzaker:bk:1992}.

To speak of an equilibrium (aka stationary) distribution for $S$, we need a unique pmf on $\mathbb{S}$, $\pi$, such that $\pi$ the marginal distribution of $S_t$ for all ``random'' $t$ (in a sense to be explained).  Formally, $\pi$ is said to be the \emph{equilibrium distribution} of $S$ if, for all times $t$ and intervals $\delta>0$, and states $s$, $\Pr(S_t=s)=\pi_s$ implies $\Pr(s_{t+\delta}=s)=\pi_s$.  In other words, if we find that $S_t$ has marginal distribution $\pi$ at one point in time, it must also have marginal distribution $\pi$ at future times.  This condition (which is often written in terms of a time 0, instead of time $t$) is usually thought of in terms of initialization: if we manually initialize a CTMC with its equilibrium distribution, it should remain in equilibrium.  Here, however, we instead take this as a characterization of the notion of observing the system at a ``random time:'' if we sample the system in a manner that does not bias our observations with respect to $\pi$, then we will observe its future states to preserve this distribution.  By contrast, if we were to choose $t$ such that the system state were badly biased (``out of equilibrium''), there would exist future times for which bias would also be preserved.  Such state selection is the true basis for the illusion of an ``arrow of time'' in reversible systems,\footnote{E.g., spontaneous deflation of a balloon is rather faster than the time required for it to spontaneously reinflate.  But this seeming asymmetry is driven by the fact that we \emph{selected} the balloon when it was in a very rare conformation, and the waiting time to observe such a rare conformation is very long. The ``arrow'' is in our selection process, not the system itself.  This is merely the regression effect, much disguised.} and occasions much mischief.  We also note from the outset the trivial but important fact that if $S$ has an equilibrium distribution then so must $G(S)$: defining $H$ to be the mapping from $\mathbb{S}$ to $\mathbb{G}$ such that $H_{ij}=1$ if $G(i)=j$ (else 0), then the equilibrium distribution $\mu$ on $\mathbb{G}$ is $H \pi$ (where we notationally treat pmfs as column vectors and $H$ as a matrix). 

It is convenient for our purposes to define a transition probability function $P$, for $S$, such that $P_{ij}(t)=\Pr(S_{\delta+t}=j|S_t=i)$ for arbitrary $\delta$.  (Note that since we have designed our rate structure to be time homogeneous, our choice of $\delta$ is indeed arbitrary.)  $S$ is said to be \emph{irreducible} iff the graph formed by the dichotomized adjacency structure $R>0$ is strongly connected.  ($T$ being strongly connected is thus a sufficient condition for irreducibility.)  Iff $S$ is irreducible, then $P_{ij}(t)>0$ for all $i,j\in \mathbb{S}$, and times $t$.  Clearly, these conditions are also inherited by $G(S)$, so an irreducible CTMC on $\mathbb{S}$ is also an irreducible CTMC on $\mathbb{G}$.

\paragraph{Irreducible Finite Chains:} The simplest sufficient conditions for equilibrium are that (1) $S$ is irreducible, and (2) $|\mathbb{S}|<\infty$.  In this case, there exists a unique $\pi$ such that (i) $\lim_{t\to \infty}P_{ij}(t)=\pi_j$ for all $i,j\in \mathbb{S}$, and (ii) if $\Pr(S_t=s)=\pi_s$ for all $s\in \mathbb{S}$, then $\Pr(S_{t+\delta}=s')=\pi_{s'}$ for all $s'\in\mathbb{S},$ $t$, $\delta>0$.

\paragraph{Co-kernel of $R$:} Because $\pi$ must be a fixed point, we also have the condition that $\pi$ is an equilibrium distribution iff $\pi^T R = 0$ (where 0 is understood here to be the zero-vector).  This is equivalent, in intuitive terms, to the statement that $\pi$ is an equilibrium if $\pi$ makes the total incoming flux for any give state equal to its corresponding out-flux.  The solutions to this equation form the cokernel or left nullspace of $R$; the existence of such a solution is thus a necessary and sufficient condition for a stationary distribution.  If $S$ is also irreducible, then any such solution is unique.  These conditions hold regardless of the size of $\mathbb{S}$.  Another potentially useful result (which also stems from the ergodic theorem) is that if there does \emph{not} exist a stationary distribution $\pi$, then $P_{ss'}(t)\to 0$ as $t\to\infty$, for all $s,s'$.  Showing that there exist distinct states that communicate over infinite time thus contradicts the absence of an equilibrium, albeit without providing one.

\paragraph{Relationship with Embedded MCs:} Any CTMC can obviously be associated with a discrete-time Markov chain constructed from its state transitions (its embedded Markov chain).  Such a chain has state space $\mathbb{S}$ and transition probabilities $\tilde{P}_{ss'}=R_{ss'}/u_s$, where $u_s=\left(\sum_{s''\in\mathbb{S}} R_{ss''}\right)$. If $S$ is irreducible, then so is its embedded chain, and if $\tilde{\pi}$ is the equilibrium pmf of the embedded chain, then $\pi_s=(\tilde{\pi}_s/u_s)/\sum_{s'\in\mathbb{S}}\tilde{\pi}_s/u_s$.  This provides an alternative route to constructing a target distribution for $S$ by starting with a pseudo-time Markov process and adding duration information to obtain the desired target.

\subsection{Known Examples}

Although there exists an unlimited number of continuous time stochastic processes that lead to a given distribution, few are currently well-characterized.  Here, we describe all processes in the literature known to lead to general classes of random graph distributions with known ERGM forms, as well as some examples that do not appear to have previously been proposed.  In each case, we describe the process itself, and demonstrate that the process does lead to the specified equilibrium (generally by the co-kernel method described above).  In cases where such convergence is already known, examination of the inbound and outbound flux to an arbitrary state can nevertheless provide other insights into model properties; for novel processes, such demonstrations are obviously necessary. Key results from this section are summarized in tabular form in Sec.~\ref{sec_disc}.

\subsubsection{Dyad State CTMCs}

Numerous models (going back to the 1950s, at least) have been put forward for networks without dyadic dependence, that lead to homogeneous or inhomogeneous Bernoulli or Categorical graphs.  An early example is \citet{katz.proctor:p:1959}, who estimate transition probabilities between mutual, asymmetric, and null dyad states; though they do not flesh out their model in full detail, it is notable in including inhomogeneity by a vertex attribute (gender).  \citet{holland.leinhardt:jms:1977} describe time-homogeneous CTMCs on the set of order-$N$ digraphs, with the transition rates being an arbitrary function of the graph state.  Although they explicitly describe the inclusion of effects for e.g. degree and triadic statistics, the forms they describe have no simple equilibrium, and they do not attempt to obtain one (merely noting that such models can be specified).  Instead, they consider only the special case of a homogeneous dyadic process with reciprocity effects.  This same dyad process was studied by \citet{wasserman:sm:1980}, who provided a more rigorous treatment.  Special case constructions were also used by \citet{snijders.vanduijn:ch:1997}, who obtained dyad state CTMCs as special cases of the competing rate SAOMs (discussed further below). Because all Bernoulli/Categorical graphs are trivial to write in ERGM form, any model of this kind has an ERGM representation.  Analyses of these processes exploit the fact that dyads can be written as either two or four-state systems (depending on directedness), and in a dyadic independence model each dyad can simply be treated as its own system with its own dynamics; due to the small number of states, this can be treated explicitly \citep[see, e.g.][among others]{wasserman:sm:1980}.  

Because these families do not give rise to very general purpose data models - and since they arise as special cases of the other processes described here - we do not consider them in further detail here.  However, we do observe that dyad state models may sometimes have useful applications in special cases.  For instance, \citet{carnegie.et.al:jcgs:2015} use models of this kind as a building block for developing equilibrium approximations to constant dissolution rate STERGMs.  It is also plausible that inhomogeneous extensions of such processes may be useful when dynamics are driven primarily by exogenous factors.

\subsubsection{The Contact Formation Process}

A very simple example of a related class of processes with known limits (but with a more complex state space) is the contact formation process (CFP) \citep{butts:jms:2019}.  Introduced to provide a mechanistic foundation for the sparse graph reference measure of Krivitsky and colleagues (\citeyear{krivitsky.et.al:statm:2011}) (and subsequently extended \citep{butts:jms:2020a} to reproduce reference measures with constant reciprocity \citep{krivitsky.kolaczyk:ss:2015} and power law mean degree scaling \citep{butts.almquist:jms:2015}), the CFP is intended as a baseline model rather than a general specification; thus, it is by design quite simple, and leads to a limited set of equilibrium ERGMs.  We describe it here because it provides a rare example of a graph process with known equilibrium for which $\mathbb{S}\neq\mathbb{G}$.

The base CFP involves a set of $N$ vertices, $V$, each of which resides in one of $M$ locations (``foci,'' per \citet{feld:ajs:1981}).  The state space of the CFP is thus $\mathbb{G} \times \{1,\ldots,M\}^N$, unlike more typical models whose state space is synomymous with the graph support.  Vertices migrate at random between foci, with migration occurring in continuous time with constant rate $r_m$ and destinations chosen uniformly at random.  Edge variables whose endpoints reside in the same locations produce \emph{formation events} at constant rate $r_f$, and all edge variables produce \emph{dissolution events} at rate $r_d$.  In the contact formation process with reciprocity (CFPR), formation events are also produced for $(i,j)$ edge variables for which the $(j,i)$ edge is present, regardless of co-location.  An $(i,j)$ edge is then said to be present at time $t$ when the most recent $(i,j)$ event is a formation event, and such an edge is said to be absent at time $t$ when the most recent event is a dissolution event.  (Although introduced as a formal convenience, formation and dissolution events can be rationalized as reflecting opportunities or situations that might cause a relationship to form or to break.  Obviously, such opportunities only act when circumstances allow, e.g. if a tie is already present, then an occasion that would have led to the formation of said tie has no effect.)

The behavior of the CFP/R depends critically on the migration rate, $r_m$.  As $r_m/\min(r_f,r_d)\to 0$, the limiting distribution of $G$ approaches a random union of Bernoulli (respectively, Categorical) graphs, with graph memberships being distributed Categorical$(M)$.  The case of greater interest is the ``fast mixing'' regime in which $r_m/\max(r_f,r_d)\to \infty$.  In this limit, the migration process is ``blurred out'' of the graph structure, but still affects the resulting graph distribution.  To obtain useful behavior, we replace the unobserved $M$ by an assumed scaling with $N$ (intuitively, a statement about how the population spreads out as network size changes).  For the important case of constant population density ($M\propto N$), we obtain for the CFP an ERGM class with sufficient statistic $w_e$ (the edge count) and reference measure $h(g)=N^{-w_e(g)}$, i.e. the Krivitsky reference.  For the CFPR, the corresponding class has $w_e,w_m$ (edge and mutual count) as sufficient statistics, and reference measure $h(g)=N^{w_m(g)-w_e(g)}$ (i.e., the \citet{krivitsky.kolaczyk:ss:2015} measure).  Alternative choices of $M$-scaling lead to different mean degree scaling.  In particular, choosing $M \propto N^{1-\gamma}$ (hence $h(g)=N^{(1-\gamma)(w_m(g)-w_e(g))}$ for the CFPR) leads to mean degree that scales as $N^\gamma$.  This allows regeneration of the power-law degree scaling measure of \citet{butts.almquist:jms:2015}, which takes both the conventional (counting measure) and Krivitsky reference as special cases.

The CFP/R illustrates how unmodeled degrees of freedom (here, migration between foci) can alter the equilibrium graph distribution; in this case, the mean degree and/or reciprocity are impacted by restrictions on the opportunity to form ties.  As these are entropic effects, deriving them from first principles clarifies that they properly ``belong'' in the reference measure, as opposed to functioning as offsets to the edge (and, for the CFPR, mutuality) parameters (as previously supposed).  Although they can be implemented in this latter way in constant-temperature models, Eq.~\ref{eq_netham} shows that this leads to models with different temperature scaling.  Even in social systems for which ``temperature'' is not currently well-understood, being able to separate opportunity-based drivers of structure from behavioral ones is a fundamental sociological objective \citep{mayhew:jms:1984b}, and we can thus obtain greater clarity from the distinction.  Finally, the CFP/R provides us with a mechanistic account of \emph{why} mean degree/reciprocity scaling may occur, and thus where we may expect to see deviations from the typical pattern.  On the other hand, the CFP/R is by design a very limited, baseline processes, whose analysis is difficult to extend.  While it seems likely that more elaborate variants can be constructed, it not obvious that they will be suitable for deriving general-purpose ERGM families.

\subsubsection{Competing Rate SAOMs}

Plausibly the first - and one of the most interesting - families of CTMCs leading to general ERGM equilibria was proposed by \citet{snijders:sm:2001} as a special case of the SAOMs.\footnote{This formulation itself appears earlier \citep{snijders.vanduijn:ch:1997}, but is not discussed as a general ERGM generating process.}  Although he does not give this specific case a name, we here refer to the models as ``competing rate'' SAOMs (since the system can be viewed as a set of potential actions ``racing'' to be next to be chosen based on their utility gain to the respective actor).  They may be summarized as follows.  We begin by taking $\mathbb{G}$ to be the set of order-$N$ digraphs, and setting $\mathbb{S}=\mathbb{G}$; $S$ is hence synonymous with a process that operates directly on the order-$N$ digraphs, and in what follows we simply refer directly to $G_t$ rather than $G(S_t)$.  Actions within the model represent single-edge changes by individual agents (who are presumed to control the state of their outgoing edges).  To represent such changes, let $g^c_{ij}$ be the graph $g$ with the state of the $i,j$ edge ``toggled'' (i.e., the $(i,j)$ edge added if absent, or otherwise removed).  Opportunities for individuals to act occur as Poisson-like events with piecewise constant hazards, such that the hazard of agent $i$ being able to act at time $t$ is
\begin{equation}
\lambda_i(G_t) = \sum_{j\in V\setminus i} \exp\left[q((G_t)^c_{ij})\right].
\end{equation}  
The potential function $q(g_t)$ is taken to encode changes in actor utilities (making this a potential game \citep{young:bk:1998}), and in particular it is assumed that if actor $i$ is able to act, he or she will choose to alter his or her tie to actor $j$ by the multinomial logit
\begin{equation}
\Pr(i\leadsto j|\theta,G_t) = \frac{\exp(q((G_t)^c_{ij}))}{\sum_{k\in V\setminus i}\exp(q((G_t)^c_{ik}))},
\end{equation}
where $i\leadsto j$ denotes the choice of $i$ to toggle the $(i,j)$ edge.  The network then evolves by the dual process of actor activations and subsequent toggles.

From this, we can immediately infer that $T$ is the set of Hamming adjacencies in $\mathbb{G}$ (since graph $g$ can instantaneously transition to $g'$ iff they differ by one edge).  To obtain $R$, begin by denoting the single changed edge for an $a,b$ transition by $(i,j)$.  This transition can only occur if (1) $i$ gets to choose, and (2) $i$ chooses to toggle the $(i,j)$ edge.  This occurs with rate 
\begin{align}
R_{ab} &= \lambda_i(a) \Pr(i\leadsto j|\theta,a) \\
&= \frac{\sum_{k\in V\setminus i} \exp\left[q(a^c_{ik})\right] \exp(q(a^c_{ij}))}{\sum_{k\in V\setminus i}\exp(q(a^c_{ik}))}\\
&=\exp\left[q(a^c_{ij})\right]\\
&=\exp\left[q(b)\right].
\end{align}

Since the set of transitions is connected, and the state space is finite, the competing rate SAOM has an equilibrium distribution.  We can verify that the equilibrium is $\pi_a = \exp(q(a))/Z$ by inspecting the rate structure.  Per Section~\ref{sec_notation}, let $\mathcal{H}(a)$ be the set of Hamming neighbors of $a$.  The incoming flux to state $a\in\mathbb{G}$ under the hypothesized equilibrium is then
\begin{align}
\sum_{b \in \mathcal{H}(a)} \pi_b R_{ba} &= \sum_{b \in \mathcal{H}(a)} \left[\exp(q(b))/Z\right] \exp\left[q(a)\right]\\
&= \exp(q(a))/Z \sum_{b \in \mathcal{H}(a)} \exp(q(b))\\
&=\pi_a \sum_{b \in \mathcal{H}(a)} R_{ab}\\
&=-\pi_a R_{aa}.
\end{align}
Since this is true for all $a\in\mathbb{G}$, it then follows that $\pi^T R=0$, and $\pi$ is the equilibrium distribution of the competing rate SAOM process.

As alluded to above, one feature of this process is that it implies that actors move faster when they ``have a chance at a good thing,'' and move more slowly when not surrounded by favorable options.  This property is the precisely that implied by the agent-based interpretation of the ``vanilla'' REM specification \citep{butts:sm:2017}, and indeed this process provides perhaps the most natural example of the connection between REMs, SAOMs, and ERGMs.  Whether this is a desirable or realistic property, of course, depends on one's system of interest; it is a fairly natural model for e.g. freely acting agents facing no exogenous constraints on their choices, but not e.g. for systems in which change rates may be limited by exogenous factors.  In particular, note that there is no upper limit on the pace of change in the competing rate SAOM, and the system will transition to favorable states exponentially fast.  Moreover, since transitions depend only on the potential of the target state, a competing rate SAOM will be just as fast to move to a favorable target state from another favorable state as from an unfavorable state.  This can result in rapid cycling between high-potential states.  Below, we will see examples of other model classes that do not have this property.

We comment in passing that the rate specification of the competing rate SAOM is only one of many possible choices, and indeed it is not widely used in practice.  Another, more common specification retains the SAOM choice process, but sets the per-actor event rates to some constant, $A$.  (This is also commonly used in DyNAMs.)  To our knowledge, the equilibrium for this family in the general case remains an open problem.

\subsubsection{Longitudinal ERGMs}

Koskinen, Snijders, and colleagues \citep{koskinen.snijders:jspi:2007,koskinen.et.al:ns:2015} proposed a general family of CTMC processes with specified ERGM distributions (inspired by prior work on SAOMs, and on quasi-time Gibbs samplers), which they dubbed ``longitudinal ERGMs'' (LERGMs).  Subsequently, \citet{grazioli.et.al:jpcB:2019} showed that a formally equivalent process could be derived from physical principles.  Since this last provides some additional motivation for the LERGM framework, our exposition here combines ideas from both the statistical and physical points of view. 

As with the competing rate SAOMs, the LERGMs take $\mathbb{G}$ to be the set of order-$N$ (di)graphs, and set $\mathbb{S}=\mathbb{G}$.  We take $T$ to be the set of Hamming adjacencies on $\mathbb{G}$ (i.e., $S$ evolves through single-edge changes), and further take $R_{ab}=0$ iff $(a,b) \not\in T$.

We proceed by starting with the target distribution, and working backwards.  We first observe that $\mathbb{G}$ is finite and Hamming connected, so it follows that the LERGM process has an equilibrium distribution.  Let us posit that $G$ has an equilibrium ERGM distribution, such that $\Pr(G_t=g|\theta)= \exp(q(g))/Z$.  Now, consider two Hamming-adjacent states $a$ and $b \in \mathbb{G}$.  The conditional probability of finding the system in state $b$, given that it is in either $a$ or $b$, is
\begin{align}
\frac{\Pr(G_t=b|\theta)}{\Pr(G_t=a|\theta)+\Pr(G_t=b|\theta)} &= \frac{\exp\left[q(b)\right]/Z}{\exp\left[q(a)\right]/Z + \exp\left[q(b)\right]/Z} \\
 &= \frac{1}{1+\exp\left[ q(a)-q(b)\right]} \label{eq_deltakin}
\end{align}
where $q(b)-q(a)$ is the change in the ERGM potential in going from state $a$ to state $b$.   This ``change'' interpretation suggests the notion of taking the transition rate to be based on Eq.~\ref{eq_deltakin}, i.e.
\begin{equation}
R_{ab} = \frac{A}{1+\exp\left[q(a)-q(b)\right]}, \label{eq_kinrate}
\end{equation}
where $A$ is a free parameter (the rate constant) with units of inverse time, and as usual $R_{aa}=-\sum_{b\neq a}R_{ab}$.

With $R$ defined per the above, now take $\pi$ to be the target ERGM distribution and consider $\pi^T R$.  Without loss of generality, let us focus on some state $a$, and let $\mathcal{H}$ denote the Hamming neighborhood of a given state.  The incoming flux into $a$ is equal to 
\begin{align}
\sum_{b\in \mathcal{H}(a)} \pi_b R_{ba} &= \frac{A}{Z} \sum_{b\in \mathcal{H}(a)} \exp\left[q(b)\right]  \frac{\exp\left[q(a)\right]}{\exp\left[q(a)\right] + \exp\left[q(b)\right] }\\
&= \frac{A \exp\left[q(a)\right]}{Z} \sum_{b\in \mathcal{H}(a)}  \frac{\exp\left[q(b)\right]}{\exp\left[q(a)\right] + \exp\left[q(b)\right] }\\
&= \pi_a \sum_{b\in\mathcal{H}(a)} R_{ab}\\
&= - \pi_a R_{aa},
\end{align} 
Since this relationship holds for all $a$, it thus follows that $\pi^T R =0$, and the equilibrium of the LERGM process is ERGM$(\theta,h)$ distributed.

The physical motivation for the LERGM process stems from the observation that, in a physical setting, Eq.~\ref{eq_kinrate} approximates the well-known Arrhenius law of chemical kinetics.  In this interpretation, each potential transition is viewed as a competing ``reaction'' (literal or metaphorical), and the resulting network arises from their joint kinetics.  In this regard, $q(b)-q(a)=\theta^T(w(b)-w(a)) + (\log h(b)-\log h(a))$ corresponds to the $-1/k_B T$-scaled free energy change associated with a specific edge toggle; thus, energy-increasing moves are unfavorable (with the rates becoming approximately $A\exp(-\Delta E/k_BT)$ when the free energy change $\Delta E$ is large), while energy-decreasing moves are favorable (with a rate that approaches $A$).  The constant $A$ corresponds to the \emph{collision rate,} a microscopic upper bound on the pace of change in the system.  This highlights the fact that the LERGM process is thus dynamically asymmetric: transitions can be arbitrarily slow, but there is an upper bound on how rapid they can be (which is approached exponentially fast for favorable transitions).  In a social context, a parallel interpretation is that the system experiences myriad randomly occurring perturbations that could lead absent ties to form, or existing ties to dissolve.  These vary in strength, but independently impact all relationships (arriving at rate $A$).  The chance of any given relationship being altered depends on the change in potential associated with the corresponding toggle: the more unfavorable the change, the stronger a perturbation must be for it to occur (and the lower the rate).  By contrast, more favorable transitions are increasingly likely to occur; however, a perturbation is still needed to activate the change, eventually leading to $A$ as the limiting rate.

Another feature of the LERGM process is that the ``dwell time'' in a given state depends on the gap in the ERGM potential (i.e., log equilibrium probability) between the state and its neighbors.  A state that is substantially more favorable than its neighbors will have very long occupancies (conditional on entry), compared to one that is less favorable than its neighbors.  Likewise, independence of rates across state-dyads means that $R_{ab}$ is neither enhanced nor inhibited by the presence of high or low $R_{ac}$ values for some third state $c$.  Although the \emph{probability} of an $a\to b$ transition obviously falls when some $R_{ac}$ rises, the \emph{rate} of this transition is not affected.

Finally, we observe that there is a certain resemblance between the LERGM process and the updates used by the standard edge-toggle Gibbs sampler \citep[a point emphasized by][]{koskinen.et.al:ns:2015}; this is because both formulations begin by considering the conditional distribution of a state pair, though this distribution is not used in exactly the same ways.  Note in particular that \emph{all} transitions are competing in the LERGM process, and not simply those within a single dyad, and transition opportunities do not occur at random.  The embedded Markov chain induced by the LERGM process is not therefore a Gibbs sampler, despite their superficial similarity.

\subsubsection{The Change Inhibition Process}

This process has not to our knowledge appeared in the prior literature, but is an obvious counterpart to the LERGM process.  As with the LERGM development, take $\mathbb{S}=\mathbb{G}$ for $\mathbb{G}$ being the set of order-$N$ graphs (or digraphs), with $T$ corresponding to Hamming adjacency and $R_{ss'}>0$ iff $(s,s')\in T$.  Let us presume the presence of a (social or otherwise) potential, $q$, such that $q:\mathbb{G}\to\mathbb{R}$.  We imagine that a transition from graph state $a$ to state $b$ occurs at some constant rate (i.e., ``at random'') when $q(b)\ge q(a)$, but that downhill moves are inhibited; specifically, given states $a,b\in\mathbb{G}$, we posit
\begin{equation}
R_{ab} = A \min(1, \exp(q(b)-q(a)).
\end{equation}

We refer to this as the ``change inhibition'' process, because its dynamics are entirely driven by inhibition of unfavorable transitions.  We now show that, like the LERGM, this process has an ERGM stationary distribution.  First, we observe that (by the same arguments as the LERGM case), this is a irreducible process on a finite state space, and hence has an equilibrium distribution.  We posit that this distribution is given by $\pi_a = \exp(q(a))/Z$.  As before, we proceed by considering the flux into an arbitrary state, $a$, using $\mathcal{N}^+$ to denote the set of neighboring states $b$ such that $q(b)\ge q(a)$, and $\mathcal{N}^-$ to denote the set of neighboring states such that $q(b)< q(a)$.  We then have
\begin{align}
\sum_{b\in\mathcal{N}(a)} \pi_b R_{ba} &= \frac{A}{Z} \sum_{b\in\mathcal{N}(a)} \exp(q(b)) \min(1,\exp(q(a)-q(b)))\\
 &= \frac{A}{Z} \sum_{b\in\mathcal{N}^+(a)} \exp(q(a)) + \frac{A}{Z} \sum_{b\in\mathcal{N}^-(a)} \exp(q(b))\\
 &= \frac{\exp(q(a))}{Z} \left[ \sum_{b\in\mathcal{N}^+(a)} A + \sum_{b\in\mathcal{N}^-(a)} A\exp(q(b)-q(a)) \right]\\
 &= \pi_a \left[ \sum_{b\in\mathcal{N}^+(a)} R_{ab} + \sum_{b\in\mathcal{N}^-(a)} R_{ab} \right]\\
&= \pi_a \sum_{b\in\mathcal{N}(a)} R_{ab}\\
&= -\pi_a R_{aa},
\end{align}
which establishes that $\pi^T R=0$, and the equilibrium is as proposed.

It is immediately apparent that while the LERGM rate function resembles the transition function of a Gibbs sampler, the change inhibition process resembles the corresponding Metropolis algorithm.  However, it is again important to note that in this process all change events are competing with each other, and it is not equivalent to quasi-time Metropolis dynamics.  Like the LERGM, dwell times at each state are related to the gaps in favorability between the focal state and its neighbors, but in this case uphill moves are not enhanced; thus, dwell times will be longer for low-potential states with high-potential neighbors.  Also like the LERGM (but not the competing rate SAOM), the change inhibition process has a maximum rate of change (which is realized for all uphill moves).  

\subsubsection{The Differential Stability Process}

Another process that has not to our knowledge appeared in the prior literature (but that is motivated by well-known properties of CTMCs) arises from allowing changes between states to occur at random, with only the dwell time in individual states varying in a systematic manner.  Specifically, we consider a process in which the expected time spent in an arbitrary state, $a$, after entry is proportional to $\exp(q(a))$.  In this \emph{differential stability process,} some states are more favorable, or stable, than others, and the system will spend more time in such states before transitioning out of them.  However, the transitions themselves are to random Hamming neighbors.  Such a system lies at the opposite extreme from a constant rate SAOM, where aggregate event rates are held constant and transitions are driven by differences in the conditional probabilities of transitioning between competing states, and is hence interesting as a corner case; substantively, however, it may be a useful starting point for modeling ``win-stay, lose-shift'' dynamics, or network evolution that is driven primarily by exogenous shocks.  

For the differential stability process, we take $\mathbb{S}=\mathbb{G}$, and let $T$ be given by Hamming adjacency.  We define the process by specifying a rate structure, $R$, such that for Hamming-adjacent neighbors $a$ and $b$, $R_{ab}=A |\mathcal{H}(a)|^{-1} \exp(-q(a))$, with transition rates of 0 for $b \not\in \mathcal{H}(a)$.  Note that, per the above, this rate does not depend upon the destination state.  It is clearly the case that the dwell time here in state $a$ is then $-1/R_{aa} = A \exp(q(a))$, as desired, with $A\in\mathbb{R}^+$ being a pacing constant.  Since the rate structure of this system is connected on the state space, and since the state space is finite, the process has an equilibrium distribution.  We posit that this equilibrium has the form $\pi_a=\exp(q(a))/Z$.  To show that this is so, we first consider the flux into an arbitrary state $a$:
\begin{align}
\sum_{b\in \mathcal{H}(a)} \pi_b R_{ba} &= \sum_{b\in \mathcal{H}(a)} \frac{\exp(q(b))}{Z} A |\mathcal{H}(a)|^{-1} \exp(-q(b))\\
&= \frac{A}{Z|\mathcal{H}(a)|} \sum_{b\in \mathcal{H}(a)} 1\\
&=A/Z.
\end{align}
Now, consider the outgoing flux at $a$:
\begin{align}
\pi_a \sum_{b\in \mathcal{H}(a)} R_{ab} &= \frac{\exp(q(a))}{Z} \sum_{b\in \mathcal{H}(a)}  A |\mathcal{H}(a)|^{-1} \exp(-q(a))\\
&= A/Z.
\end{align}
The incoming flux equals the outgoing flux, and hence $\pi$ is indeed the equilibrium distribution of the differential stability process.

Although the property is not unique to this family, the fact that transitions are uniformly random in the differential stability process makes especially obvious the fact that inference for its parameters can be performed using only aggregate information on state occupancies, without sequential information.  This may facilitate the use of this model with fragmentary data sources, or in other settings where it is easier to estimate where the system is spending most of its time than to track specific transitions.

\subsubsection{Continuum STERGMs}

As discussed, the STERGM class of temporal ERGMs has been an important focus of work related to continuum-limit models, because of the potential for fitting to limited observational data sets.  The constant dissolution model has been central to this effort, and we treat it here separately, followed for by its constant formation counterpart.  While less trivial to fit, general STERGMs can also be extended to the continuum limit, and we close this section with a consideration of this broader family of cases.

\paragraph{Constant-dissolution Continuum STERGMs}

As noted above, \citet{carnegie.et.al:jcgs:2015} employ STERGMs with constant dissolution rates as a tool for approximate estimation of network dynamics from combined cross-sectional and durational data; they obtain exact results for the dyadic independence case, and then argue from simulation that such a STERGM will approach an approximate ERGM equilibrium whose potential is equal to the potential of the formation model, with an edge parameter adjustment for dissolution.  In an inferential setting, this parameter can be directly estimated given tie duration.  Subsequently, \citet{klumb.et.al:wp:2022} have shown that the error in this approximation declines as the length of the (discrete) STERGM time step decreases, and one does (under mild regularity conditions) approach the presumed limit as step size goes to zero.  They refer to the model under this finite but arbitrarily small step size an ``infinitesimal STERGM.''  Here, we directly consider the corresponding continuous time process, which we refer to as a ``continuum STERGM'' (CSTERGM) with constant dissolution rates.  Specifically, we here consider the case of a CSTERGM in which rates of edge formation are driven by an ERGM potential $q_f$, while dissolution is driven by a single parameter $\theta_d$ (i.e., all edges are lost at constant rate $\exp(\theta_d)$).  

We construct the continuum process as follows.  Let $S$ be a CTMC on state space $\mathbb{S}=\mathbb{G}$, with $\mathbb{G}$ being the set of all order-$N$ (di)graphs and $T$ being the Hamming adjacency on the state space.  Let $\mathcal{H}^+$ be the set of Hamming neighbors formed by adding an edge to the argument, and let $\mathcal{H}^-$ be the set of Hamming neighbors formed by removing an edge from the argument.  $S$ then follows the instantaneous rate structure
\begin{equation}
R_{ab} = \begin{cases}   
0 & \text{ if } b\not\in \mathcal{H}^+(a) \cup \mathcal{H}^-(a)\\
\exp(\theta_d) & \text{ if } b\in \mathcal{H}^-(a)\\
\exp(q_f(b)-q_f(a)) & \text{ if } b\in \mathcal{H}^+(a)\\
-\sum_{b\in\mathbb{G}\setminus a} R_{ab} & \text{ if } a=b
 \end{cases}.
\end{equation}

Clearly, $|\mathbb{S}|<\infty$, and $T$ is connected; thus, $S$ has an equilibrium distribution.  We posit that this equilibrium distribution corresponds to $\pi_a = \exp(q_f(a)-\theta_d w_e(a))/Z$.  To verify this, we consider the influx to an arbitrary state $a$:
\begin{align}
\sum_{b\in\mathbb{G}\setminus a} R_{ba} & =\sum_{b\in\mathcal{H}^+(a)} \pi_b \exp(\theta_d) + \sum_{b\in\mathcal{H}^-(a)} \pi_b \exp\left(q_f(a)-q_f(b)\right)\\
&= \sum_{b\in\mathcal{H}^+(a)} \frac{\exp(q_f(b)-\theta_d w_e(b))}{Z} \exp(\theta_d) + \sum_{b\in\mathcal{H}^-(a)} \frac{\exp(q_f(b)-\theta_d w_e(b))}{Z} \exp\left(q_f(a)-q_f(b)\right)\\
&=\frac{1}{Z} \left[\sum_{b\in\mathcal{H}^+(a)}  \exp\left(q_f(b)-\theta_d (w_e(b)-1)\right) + \sum_{b\in\mathcal{H}^-(a)}  \exp\left(q_f(a)-\theta_d w_e(b)\right)\right].\\
\intertext{Observe that, for $b\in\mathcal{H}^+(a),$ $w_e(b)=w_e(a)+1$, and for $b\in\mathcal{H}^{-}(a),$ $w_e(b)=w_e(a)-1$.  Substituting and factoring out the putative equilibrium probability of $a$ gives}
&= \frac{\exp(q_f(a)-\theta_d w_e(a))}{Z} \left[ \sum_{b\in\mathcal{H}^+(a)} \exp\left(q_f(b)-q_f(a)\right) + \sum_{b\in\mathcal{H}^-(a)} \exp(\theta_d) \right]\\
&= \pi_a \sum_{b\in\mathbb{G}\setminus a} R_{ab}\\
&=-\pi_a R_{aa}.
\end{align}
It thus follows that $\pi^T R=0$, and the continuum STERGM has the desired equilibrium.

\paragraph{Constant Formation Continuum STERGMs}

While constant dissolution STERGMs have received more attention in the literature, a very similar development is also possible for constant \emph{formation} STERGMs: processes in which edges form at a constant rate, but where their persistence is governed by a potentially complex process.  Such a model represents relational ``selection'' in its truest sense, since the key dynamics are driven by the differential survival of relationships.  That this side of the STERGM family is less well-studied than the constant dissolution side may reflect a bias in the field towards processes that create edges rather than those that retain or remove them (a broader concern raised e.g. by \citet{burt:sn:2000,burt:sn:2002}), though the constant dissolution approximation is doubtless reasonable in many settings.  In some cases, however, ties may form initially through processes that are idiosyncratic and largely exogenous, with enduring structure resulting from the maturation of some fraction of those ties into longer-term relations.  For instance, in work settings structured around \emph{ad hoc} teams convened to perform specific projects, tie formation may be driven by short-term assignments that result from shifting labor needs, rather than any preferences of the individuals involved.  Some, however, may elect to keep in touch with colleagues from those short-term assignments over the long haul, a process that may be driven by individual considerations.  Alternately (as with the differential stability process), network evolution may not involve individuals or decision processes at all, and may simply reflect differences in the stability of relationships to exogenous shocks.  

Although, to our knowledge, a continuum version of the constant formation STERGM has not appeared in the literature, we can easily define it here.  As with the constant dissolution STERGM, we take $\mathbb{S}=\mathbb{G}$, presume that $\mathbb{G}$ is the set of order-$N$ graphs, and employ a separate dissolution potential $q_d(a) = \theta_d^T w_d(a)$ and constant formation parameter $\theta_f$.  Our rate function is given by
\begin{equation}
R_{ab} = \begin{cases} 
\exp(\theta_f) & \text{if } b\in\mathcal{H}^{+}(a)\\
\exp(q_d(b)-q_d(a)) & \text{if } b\in\mathcal{H}^{-}(a)\\
-\sum_{b\in\mathcal{H}(a)}R_{ab} & \text{if } a=b\\
0 & \text{otherwise}
\end{cases},
\end{equation}
and we propose the equilibrium distribution $\pi_a=\exp\left[q_d(a)+w_e(a) \theta_f\right]/Z$, with $w_e$ as usual being the edge count.  To verify, we examine the in-flux to an arbitrary state $a$ in equilibrium,
\begin{align}
\sum_{b\in\mathcal{H}(a)} \pi_b R_{ba} &= \sum_{b\in\mathcal{H}^{-}(a)}\pi_b \exp[\theta_f] + \sum_{b\in\mathcal{H}^{+}(a)} \pi_b \exp\left[q_d(a)-q_d(b)\right]\\
&= \frac{1}{Z} \left[ \sum_{b\in\mathcal{H}^{-}(a)}\exp\left[q_d(b)+w_e(a) \theta_f\right] +  \sum_{b\in\mathcal{H}^{+}(a)} \exp\left[q_d(a)+(w_e(a)+1) \theta_f\right]\right], \\
\intertext{where we have used the relationship between the edge counts of $a$ and $b$, respectively.  Pulling out $\pi_a$ then gives us}
&= \frac{\exp\left[q_d(a)+w_e(a)\theta_f\right]}{Z} \left[ \sum_{b\in\mathcal{H}^{-}(a)} \exp\left[q_d(b)-q_d(a)\right] + \sum_{b\in\mathcal{H}^{+}(a)} \exp\left[\theta_f\right]\right]\\
&= \pi_a \sum_{b\in\mathcal{H}(a)} R_{ab}\\
&= -\pi_a R_{aa}.
\end{align}
The condition that $\pi^T R =0$ is then satisfied, and the constant formation CSTERGM has the posited equilibrium.

As with the constant dissolution CSTERGM, the constant formation CSTERGM can in principle be fit to data containing only information on cross-sectional network structure and durations, here specifically the duration of \emph{nulls} rather than edges; equivalently, edges added per unit time will suffice, provided that care is taken to control for the current density (since the total rate of edge addition scales with the number of edges that are available to add).  We do not consider this problem in greater detail here, but the approach parallels that for the constant dissolution case.

\paragraph{General Continuum STERGMs}

While the constant rate cases are of particular pragmatic interest, it should be noted that the same general approach used above can be used to provide a continuum version of STERGMs with arbitrary rate structure.  These general CSTERGMs have not, to our knowledge, been treated in previous literature.  We define them as follows.  Combining our previous two cases, let $q_f$ be a formation potential, and $q_d$ a dissolution potential.  We take $\mathbb{S}=\mathbb{G}$ on the order-$N$ (di)graphs, and let $T$ correspond to Hamming adjacency, with transition rates for states $a,b\in\mathbb{S}$ given by
\begin{equation}
R_{ab} = \begin{cases}   
0 & \text{ if } b\not\in \mathcal{H}^+(a) \cup \mathcal{H}^-(a)\\
\exp(q_d(b)-q_d(a)) & \text{ if } b\in \mathcal{H}^-(a)\\
\exp(q_f(b)-q_f(a)) & \text{ if } b\in \mathcal{H}^+(a)\\
-\sum_{b\in\mathbb{G}\setminus a} R_{ab} & \text{ if } a=b
 \end{cases}.
\end{equation}

Since the system is irreducible and has a finite state space, it has an equilibrium distribution, $\pi$.  We posit that $\pi_a = \exp(q_f(a)+q_d(a))/Z$ for $a \in \mathbb{S}$.  As usual, we verify by considering the influx to an arbitrary state, $a$:
\begin{align}
\sum_{b\in\mathcal{H}(a)} \pi_b R_{ba} &= \sum_{b\in\mathcal{H}^{-}(a)}\pi_b \exp[q_f(a)-q_f(b)] + \sum_{b\in\mathcal{H}^{+}(a)} \pi_b \exp\left[q_d(a)-q_d(b)\right]\\
&=\frac{1}{Z}\left[ \sum_{b\in\mathcal{H}^{-}(a)} \exp[q_f(a)+q_d(b)] + \sum_{b\in\mathcal{H}^{+}(a)} \exp\left[q_d(a)+q_f(b)\right] \right]\\
&=\frac{\exp\left[q_f(a)+q_d(a)\right]}{Z} \left[ \sum_{b\in\mathcal{H}^{-}(a)} \exp[q_d(b)-q_d(a)] + \sum_{b\in\mathcal{H}^{+}(a)} \exp\left[q_f(b)-q_f(a)\right] \right]\\
&= \pi_a \sum_{b\in\mathcal{H}(a)} R_{ab}\\
&=-\pi_a R_{aa}.
\end{align}
Thus, $\pi^T R=0$, and the equilibrium is as posited.

Like their discrete counterparts, the general CSTERGMs are particularly natural in settings for which the factors shaping edge formation are different from those shaping persistence or dissolution of edges.  Inference for these models, however, is more complex than for their constant formation/dissolution counterparts, due to the need to separate the two distinct potentials ($q_f$ and $q_d$) that cannot be inferred from mean durations.  We briefly comment further on this issue in Section~\ref{sec_disc}.

\subsubsection{Continuum TERGMs}

Having seen continuum forms of the STERGM family, it is reasonable to ask if there is a similar continuum extension of the general TERGMs.  The answer is affirmative, as can be appreciated by simply imposing $q=q_f=q_d$ in the CSTERGM development.  We thus state here the general form of the continuum TERGM process (CTERGM), which has not to our knowledge appeared in prior literature.

Following the same assumptions as the CSTERGMs, we take $\mathbb{S}=\mathbb{G}$ with $\mathbb{G}$ being the order-$N$ (di)graphs, with $T$ corresponding to Hamming adjacency on $\mathbb{G}$ and rate structure
\begin{equation}
R_{ab} = \begin{cases}   
0 & \text{ if } b\not\in \mathcal{H}(a)\\
\exp(q(b)-q(a)) & \text{ if } b\in \mathcal{H}(a)\\
-\sum_{b\in\mathbb{G}\setminus a} R_{ab} & \text{ if } a=b
 \end{cases}.
\end{equation}
Finiteness of the state space and irreducibility of the transition structure give us an equilibrium $\pi$, which we assert has the form $\pi_a=\exp(2q(a))/Z$.  We verify this by examining the influx to an arbitrary state, $a\in \mathbb{S}$:
\begin{align}
\sum_{b\in\mathcal{H}(a)} \pi_b R_{ba} &= \sum_{b\in\mathcal{H}(a)}\pi_b \exp[q(a)-q(b)]\\
&=\frac{1}{Z} \sum_{b\in\mathcal{H}(a)}\exp[q(a)+q(b)]\\
&=\frac{\exp[2 q(a)]}{Z} \sum_{b\in\mathcal{H}(a)}\exp[q(b)-q(a)]\\
&= \pi_a \sum_{b\in\mathcal{H}(a)} R_{ab}\\
&= - R_{aa}.
\end{align}
This establishes that $\pi^T R=0$, and $\pi$ is indeed the CTERGM equilibrium.

While the factor of two that appears in the CTERGM equilibrium may appear counterintuitive, we can appreciate from the CSTERGM case that it arises from the interaction of the flux out of $a$ from edge loss (respectively edge gain) and the flux into $a$ from neighbor edge gain (respectively, neighbor edge loss): when $a$ is a high-potential state, it both sends less flux to its neighbors and gets more flux in return.  In some cases, it may be more aesthetic to drop this factor of two from the equilibrium graph potential, and divide the log transition rate by two instead; we keep the present parameterization, however, to emphasize the relationship with the CSTERGMs.

\section{Discussion} \label{sec_disc}

As we have seen, there are many ways of defining continuous time graph processes that lead to ERGM equilibria, underscoring the point that the same cross-sectional distribution can arise from different generative processes.  However, not all such processes are equally plausible in any given setting.  Here, we briefly comment on what can be said about the general properties of the processes described above, and how they may be usefully distinguished on qualitative grounds.  We also note some potential insights from these processes regarding network evolution more generally, as well as some observations for practical use in applied settings.

\subsection{Summary of General ERGM-generating Processes}

Leaving aside special case models, the eight families of ERGM-generating processes are summarized in Table~\ref{tab_eq}.  (Note that, since the time scale on which the model is defined is arbitrary, all rates are defined up to a similarity transform.  For clarity, we leave the fundamental rate constants, $A$, when used in model definition.)  The diversity of formulations seen in our above development is on display here.  However, some general patterns are evident.  Transitions can be governed by the absolute potential of the target state ($q(b)$, in an $a \to b$ transition), the absolute potential of the originating state, or the difference in potentials; this reflects three broad classes of dynamics, where transitions are governed by (respectively) the favorability of the target, the unfavorability of the source, or the gain in favorability associated with transitioning from source to target.  One or another of these may be more plausible in particular settings, suggesting a specific choice of model family.  

\begin{table}
\centering
{
\setlength{\extrarowheight}{1.25em}
{\resizebox{\textwidth}{!}{%
\begin{tabular}{p{0.15\textwidth}ccc} \hline\hline
Class & Event Rate ($a \to b$) & Exit Rate ($a\to$ any) &  Equilibrium \\ \hline
Competing Rate SAOM & $\exp\left[q(b)\right]$ & $\sum_{b \in \mathcal{H}(a)} \exp\left[q(b)\right]$ & $\exp(q(a))/Z$\\
LERGM & $A\left[1+\exp\left[q(a)-q(b)\right]\right]^{-1}$ & $A \sum_{b\in\mathcal{H}(a)} \left[1+\exp\left[q(a)-q(b)\right]\right]^{-1}$ &  $\exp(q(a))/Z$\\
Change \break Inhibition & $A \min(1, \exp(q(b)-q(a))$ &  $\begin{matrix} A |\mathcal{N}^{+}(a)| + \\[-10pt] A\exp[-q(a)] \sum_{b\in\mathcal{N}^-(a)} \exp[q(b)] \end{matrix}$ & $\exp(q(a))/Z$ \\
Differential Stability & $A|\mathcal{H}(a)|^{-1} \exp(-q(a))$ & $A\exp(-q(a))$ & $\exp(q(a))/Z$ \\
Const. Diss. CSTERGM & $\begin{matrix} I(b\in \mathcal{H}^{-}(a))\exp\left[\theta_d\right] + \\[-10pt] I(b\in \mathcal{H}^{+}(a)) \exp(q_f(b)-q_f(a)) \end{matrix}$ & $\begin{matrix} w_e(a) \exp[\theta_d] + \\[-10pt] \exp[-q_f(a)] \sum_{b\in \mathcal{H}^{+}(a)} \exp\left[q_f(b)\right] \end{matrix}$  & $\exp(q_f(a)-\theta_d w_e(a))/Z$\\
Const. Form. CSTERGM & $\begin{matrix} I(b\in \mathcal{H}^{+}(a))\exp\left[\theta_f\right] + \\[-10pt] I(b\in \mathcal{H}^{-}(a)) \exp\left[q_d(b)-q_d(a)\right] \end{matrix} $ & $\begin{matrix} (M^*-w_e(a)) \exp[\theta_f] + \\[-10pt] \exp[-q_d(a)]\sum_{b\in \mathcal{H}^{-}(a)} \exp\left[q_d(b)\right] \end{matrix} $  & $\exp(q_d(a)+\theta_f w_e(a))/Z$\\
General CSTERGM & $\begin{matrix} I(b\in \mathcal{H}^{+}(a))\exp\left[q_f(b)-q_f(a)\right] + \\[-10pt] I(b\in \mathcal{H}^{-}(a)) \exp\left[q_d(b)-q_d(a)\right] \end{matrix} $ & $\begin{matrix} \exp[-q_f(a)] \sum_{b\in \mathcal{H}^{+}(a)} \exp\left[q_f(b))\right] + \\[-10pt] \exp[-q_d(a)] \sum_{b\in \mathcal{H}^{-}(a)} \exp\left[q_d(b)\right] \end{matrix} $  & $\exp(q_d(a)+q_f(a))/Z$\\
CTERGM & $\exp\left[q(b)-q(a)\right]$ & $\exp[-q(a)] \sum_{b\in \mathcal{H}(a)} \exp\left[q(b)\right]$  & $\exp(2q(a))/Z$\\
\hline\hline
\end{tabular}%
}}}
\caption{\label{tab_eq} Continuous time graph processes with known equilibria.  $A\in \mathbb{R}^+$; $q(a)=\theta^T w(a) + \log h(a)$, with $q_f,q_d$ indicating formation/dissolution potentials; $w_e$ is the edge statistic; $\mathcal{N}^{+}/\mathcal{N}^{-}$ indicate Hamming neighbors of higher/lower potential, $\mathcal{H}^{+}/\mathcal{H}^{-}$ indicate Hamming neighbors formed by edge addition/deletion; and $M^*$ is the maximum edge count.  All rates defined up to a homogeneous scale transformation.}
\end{table}

Going further, we may characterize the differences between models in terms of a number of basic properties, as shown in Table~\ref{tab_prop}.  First, we observe that two families (the LERGMs and the Change Inhibition process) have transitions rates that are bounded above by a constant - there is, for these families, a fundamental ``speed limit'' on the pace of network change, potentially reflecting unmodeled micro-level mechanisms that depend upon some other temporal process (e.g., ``collisions'' or ``encounters'') whose pace is not controlled by network structure.  For the other families, there is in principle no limit on how rapidly transitions can occur.  This is sensible where edges can evolve by near-instantaneous processes (e.g., one actor deciding in rapid succession to break off ties to each member of a group), though in other cases care may be needed in empirical settings to ensure that transition rates do not become unrealistically large.  We have also seen that families vary in whether these rates are fundamentally driven by target potential (Competing Rate SAOMs), source potential (Differential Stability), or potential differences (all others).  The target-focus of the SAOM arises from its utility-theoretic motivation, as actors' decisions within the model are assumed to be based on the attractiveness of the target state, and not on a comparison with their existing state (such comparisons being characteristic of non-rational choice models such as Prospect Theory \citep{kahneman.tversky:e:1979}).  By turns, an exclusive focus on the stability of the current state is the defining condition of the Differential Stability process.  It is interesting that all other processes so far proposed operate instead on \emph{differences} in potentials, such that dynamics slow down as the graph process enters a ``flatter'' region of the potential surface.  The motivation for difference-driven dynamics in the LERGM was originally by appeal to a conditional probability argument, although as we have seen this can also be motivated by the notion that formation and dissolution processes are ``competing'' with each other even within a dyad and that observed change rates are the result of that competition.  For Change Inhibition, the motivation is direct: we posit that the system ``resists'' downhill moves \emph{per se,} which implies a comparative structure.  Finally, the use of potential differences in the continuum (S)TERGMs is necessary for constant rate models to lead to desired behavior (i.e., to serve as continuum limits in the manner proposed e.g by \citet{carnegie.et.al:jcgs:2015} and \citet{klumb.et.al:wp:2022}).  These differences underscore the point that distinct generative mechanisms can lead to quite different dynamics.

As a next point of contrast, we see that models differ in how rates are varied across the state space.  Most immediately, the CSTERGM families are defined separably, so that formation and dissolution rates are distinct; in the general case, this feature is not shared by the other families (though some may coincide, as with the trivial case of a general CSTERGM with $q_f=q_d$ and the CTERGMs).  Less trivially, we observe that processes vary in whether transition rates vary across (are ``sensitive to'') neighboring states of higher or lower potential (respectively).  Although most processes considered here have higher rates of transition to states with much higher potential than those with only marginally higher potential, this is not true for Change Inhibition (which is indifferent between uphill moves) and Differential Stability (which is indifferent to \emph{all} moves).  This is even more true for sensitivity to variation in potential loss, with only the Differential Stability process being indifferent to transitions to states with slightly versus substantially lower potential.  These distinctions lead to very different patterns of network evolution.  Finally, we distinguish models that are sensitive to differences in potential among neighboring states formed by adding edges (members of $\mathcal{H}^{+}$) versus those formed by removing edges (members of $\mathcal{H}^{-}$).  Other than the Differential Stability process (which, as noted, is insensitive to all neighbor properties), these distinctions are characteristic of the constant rate CSTERGMs, which respectively treat all dissolution transitions or formation transitions equivalently.  Taken together, these distinctions separate all eight families, with each having a distinctive signature.  We observe that there are some signatures that do not have a corresponding model within this set, potentially suggesting opportunities for defining additional families.

\begin{table}
\centering
{\resizebox{\textwidth}{!}{%
\begin{tabular}{lccccccc} \hline\hline
Class & Max $R_{ab}$ & Driving Potential & Separable &  $\mathcal{N}^{+}$ Sensitive? & $\mathcal{N}^{-}$ Sensitive? & $\mathcal{H}^{+}$ Sensitive? & $\mathcal{H}^{-}$ Sensitive? \\ \hline
Competing Rate \newline SAOM & $\infty$ & Target & No & Yes & Yes & Yes & Yes\\
LERGM & $A$ & Difference & No & Yes & Yes & Yes & Yes \\
Change \break Inhibition & $A$ & Difference & No & No & Yes & Yes & Yes \\
Differential Stability & $\infty$ & Source & No & No & No & No & No\\
Const. Diss. CSTERGM & $\infty$ & Difference & Yes & Yes & Yes & Yes & No\\
Const. Form. CSTERGM & $\infty$ & Difference & Yes & Yes & Yes & No & Yes\\
General CSTERGM & $\infty$ & Difference & Yes & Yes & Yes & Yes & Yes\\
CTERGM & $\infty$ & Difference & No & Yes & Yes & Yes & Yes\\
\hline\hline
\end{tabular}%
}}
\caption{\label{tab_prop} Qualitative properties of model classes; graph processes vary in whether change rates are bounded (maximum $R_{ab}$), the nature of their driving potential, formation/dissolution separability, and in whether their rates vary across graphs with higher or lower potential ($\mathcal{N}^{+}$/$\mathcal{N}^{-}$ sensitivity) and/or for edge formation versus dissolution ($\mathcal{H}^{+}$/$\mathcal{H}^{-}$ sensitivity).}
\end{table}

\subsection{Insights for Network Evolution}

Considering network evolution in its most general terms provides a number of substantive insights.  First, we observe that complex structure does not imply a selective transition process: timing alone is sufficient to generate arbitrarily rich graph structure.  One cannot thus automatically assume that a focal system is driven by dynamics that seek to walk uphill on the potential (though this may be a good assumption or other reasons).  Further, this observation reminds us that factors that alter the pace of change within a social network may have a substantial impact on long-term structure, so long as they do not act uniformly.  

Another insight is that complex structure can arise equally well from formation or dissolution processes.  Though it is often reasonable to think of structure as being driven primarily by constraints on which ties can or will form (e.g., interaction opportunities, lowered or raised barriers to tie formation), it may be that differences in selection or survival of ties are the primary structural drivers.  Or both may be involved.  As above, there may be other information that leads us to believe that one class or the other will be primary in a given setting, but the mere presence of complex structure has no bearing on this.

We can also see from the processes examined here that ERGM-generating graph processes that \emph{do} follow a potential do not need to do so symmetrically: like the Change Inhibition process, they can be more sensitive to downhill moves than uphill moves (or, conceivably, \emph{vice versa}).  It may be reasonable to assume symmetry on substantive or other grounds, but it need not exist for complex structure to result.

A further insight that can be obtained by connecting dynamics with equilibrium behavior is that many dynamic processes lead to degenerate long-run behavior (in the sense of \citet{strauss:siam:1986,handcock:ch:2003}).  Indeed, any ERGM-generating process coupled with a potential leading to a degenerate ERGM distribution will also have degenerate dynamics; the fact that such a wide range of different dynamic processes can lead to the same ERGM equilibria thus confirms that degeneracy is not a consequence of any specific choice of local dynamics.  For instance, the choice-motivated SAOMs and competing rate-motivated LERGMs can both lead to the same degenerate graph distributions as CTERGMs and Differential Stability processes.  Sometimes, as per \citet{yu.et.al:nsr:2020}, this is in fact an empirically realistic outcome; for most social networks, however, degeneracy is usually pathological.  What we can glean from the present work is that - at least for processes of the type studied here - degeneracy should be understood as arising from the graph potential rather than the processes operating on it.  While this is not unprecedented (the presence of many quasi-time MCMC processes leading to identical graph distributions, for instance, suggests the same conclusion, as do results of \citet{schweinberger:jasa:2011}), it may not always have been appreciated.  A deeper consideration of the connection between dynamic processes and their long-run behavior thus has the potential to produce more general insights about the factors that lead to the types of networks we see in the real world, as opposed to the many types of networks that \emph{could} arise, but that do not.

Finally, we note that while some processes may be especially natural in particular settings (e.g., SAOMs for unilaterally controlled edges in interpersonal networks, LERGMs in physical networks), any given graph process may admit multiple interpretations.  It is important not to assume that any one mechanistic interpretation of a graph process is the only one possible.  That said, deriving an ERGM-generating graph process from first principles can provide a strong motivation for applying it in cases where the associated assumptions are met.  Mechanistic derivations of ERGM-generating continuous time graph processes would thus seem to be an important area for further research.

\subsection{Considerations for Practical Use}

Although our focus here is on model definition and general properties, we may glean a few considerations for use of these models in practice.  First, we observe that all of the ERGM-generating graph processes studied here except for the CSTERGMs have parameters that can be estimated up to a pacing constant from cross-sectional observations.  Specifically, inferring $q$ from one or more equilibrium draws determines the behavior of the model, although quantitative rates of change obviously require that the characteristic timescale of the process (and, where applicable, the maximum change rate) be inferred from information on observed dynamics.  In the case of the constant rate CSTERGMs, one cannot identify the formation or dissolution potentials (respectively) without knowing the respective dissolution or formation rates, but these are easily estimated from duration or pacing information (as discussed by \citet{carnegie.et.al:jcgs:2015}).  Moreover, since ERGM inference can often be performed using sampled or incomplete data \citep{handcock.gile:aas:2010,koskinen.et.al:sn:2013,stivala.et.al:sn:2016,krivitsky.morris:aas:2017}, it follows that a considerable amount can be inferred about network dynamics from quite limited observations, provided that one has some \emph{a priori} basis for constraining the associated model family.  

It is also useful to observe that, given an ERGM, it is straightforward to simulate hypothetical equilibrium dynamics using these models; in such applications, one must choose the pacing constant and, for the constant rate CSTERGMs, the corresponding mean durations.  As noted above, such an approach has been used in the discrete case e.g. by \citet{morris.et.al:ajph:2009} and in the continuous case by \citet{yu.et.al:nsr:2020}, both of whom used approximate matching to available empirical observations to calibrate total change rates.  In some applications, considering dynamics relative to the characteristic timescale of the process (``phenomenological time'') may be sufficient to gain useful insights.  Although a detailed discussion of simulation procedures is beyond the scope of this paper, it should be noted that standard discrete event simulation methods can be used for all of the models considered here.  In particular, given rate structure $R$ and current state $a$, the probability that the next transition will be to state $b$ is simply $R_{ab}/\sum_{c\in \mathbb{S}\setminus a} R_{ac}$, with the time to the next event being exponentially distributed with expectation $\left[\sum_{b\in \mathbb{S}\setminus a} R_{ab}\right]^{-1}$.  For these families (all of which involve single-edge changes), this can be implemented by calculating the rates for all potential toggles from the current state, then selecting both the next toggle and the time to the next event independently.  Model-specific implementations may allow for greater efficiency in some cases (e.g., by pooling events with identical rates).  It should be noted that the presence of \emph{absolute} exponentiated graph potentials in the exit rates for the Competing Rate SAOMs and Differential Stability processes (as opposed to exponentiated potential \emph{differences}; see Table~\ref{tab_eq}) implies that these processes may in some cases exhibit dynamics with very large numbers of transition events per unit phenomenological time.  For the former process, this arises when the model shuttles very rapidly between two or more Hamming-adjacent, high-potential graphs.  For the latter process, the phenomenon arises when the model has entered a low-potential region of graph space, and rapidly switches at random between low-potential graphs.  Although these dynamics take very little phenomenological time, they can become very computationally expensive in large systems with large variation in graph potential.  Specialized algorithms for handling these cases may be needed in some circumstances; where transition rates become extremely large, it may also be prudent to consider whether the process in question is substantively reasonable.

Where detailed dynamic information is available, this can of course be exploited, allowing for out-of-equilibrium inference; see, e.g. \citet{snijders:sm:2001,koskinen.snijders:jspi:2007,butts:sm:2008,stadtfeld.et.al:sm:2017}.  Such information is generally required for general CSTERGM inference (since one cannot distinguish the formation and dissolution potentials from equilibrium information), and is also needed when seeking to distinguish among ERGM-generating processes on an empirical basis.  Although model adequacy checking for continuous time models is an area of active development (see e.g. \citet{lospinoso.snijders:mi:2019}), one obvious diagnostic is to compare observed dwell times in each state to those of Table~\ref{tab_eq}.  While more conventional adequacy checking strategies based on cross-sectional simulation \citep[e.g.][]{hunter.et.al:jss:2008} are useful for assessing whether a model can reproduce equilibrium behavior, they will not distinguish among ERGM-generating processes with the same equilibrium.  By contrast, different dynamic processes with equivalent graph equilibria generally have very different inter-event timing.


\section{Conclusion} \label{sec_conc}

The above sketches some of the precursors and known classes of stochastic processes leading to known ERGM distributions.  As observed, there are several general classes of such processes; however, it is likely that these only scratch the surface of what is possible.

It is interesting to note that, with the lone exception of the CFP/R, the frameworks studied to date all take $\mathbb{S}=\mathbb{G}$.  Although this is convenient, it limits the dynamics that are possible.  Extending the state space will necessarily add an extra entropic contribution to the ERGM potential \citep[as per][]{butts:jms:2019} - equivalently, a change of reference measure - since it will change the number of ways that each graph can be realized (i.e., it adds hidden degrees of freedom).  Since entropic effects are (broadly) related to opportunity structures, state space extension may be important for capturing contextual constraints on network dynamics.

We also observe that while dynamics are limited to Hamming transitions for the LERGM, Change Inhibition, and Differential Stability processes, this limitation is not required for deriving their equilibria (other than establishing irreducibility), and it is not essential: these processes can operate (with equivalent rate functions) on any fully connected set of transitions on $\mathbb{G}$.  Thus, simultaneous edge changes can be accommodated within existing CTMC processes, albeit at computational cost.

We conclude by observing that a major limitation on progress in this area is a dearth of high-quality dynamic data on social or other networks that is capable of discriminating among competing models.  It is hoped that advances in data collection will produce a body of observations that will put continuous time network models on a firmer empirical footing.

\bibliography{ctb}


\end{document}